\RequirePackage{fix-cm}
\documentclass[twocolumn,epjc3]{svjour3}  
\smartqed  
\RequirePackage{graphicx}
\usepackage{epsfig}
\usepackage{amsmath}
\usepackage{amsfonts}
\usepackage{amssymb}
\usepackage{graphicx}
\usepackage{colordvi}
\usepackage{color}
\newcommand{\aea}{Astron. Astrophys.}
\newcommand{\apj}{Astrophys. J.}

\newcommand{\jcap}{J. Cosmol. Astropart. Phys.}
\newcommand{\mnras}{Mon. Not. R. Astron. Soc.}
\newcommand{\prd}{Phys. Rev. D}
\newcommand{\prl}{Phys. Rev. Lett}

\journalname{Eur. Phys. J. C}
\begin{document}

\title{Modified gravity revealed along geodesic tracks
}


\author{Mariafelicia  De Laurentis\thanksref{e1,addr1,addr2,addr3,addr4}
        \and
        Ivan De Martino\thanksref{e2,addr5}
         \and
        Ruth Lazkoz\thanksref{e3,addr6} 
}

\thankstext{e1}{e-mail: mariafelicia.delaurentis@unina.it}
\thankstext{e2}{e-mail: ivan.demartino@dipc.org}
\thankstext{e3}{e-mail: ruth.lazkoz@ehu.es}

\institute{Dipartimento di Fisica "E. Pancini", Universit\'a di Napoli "Federico II", Compl. Univ. di Monte S. Angelo, Edificio G, Via Cinthia, I-80126, Napoli, Italy
    \label{addr1}
           \and
            INFN Sez. di Napoli, Compl. Univ. di Monte S. Angelo, Edificio G, Via Cinthia, I-80126, Napoli, Italy
            \label{addr2}
            \and
             Institute for Theoretical Physics, Goethe University, Max-von-Laue-Str.~1, D-60438 Frankfurt, Germany
             \label{addr3}
             \and
           Lab.Theor.Cosmology,Tomsk State University of Control Systems and Radioelectronics(TUSUR), 634050 Tomsk, Russia
           \label{addr4}
           \and
Donostia International Physics Center (DIPC), 20018 Donostia-San Sebastian (Gipuzkoa)
Spain.\label{addr5} 
\and
       Department of Theoretical Physics and History of Science,  University of the Basque Country UPV/EHU, 
Faculty of Science and Technology, Barrio Sarriena s/n, 48940 Leioa, Spain
          \label{addr6}         
}

\date{Received: date / Accepted: date}

\maketitle

\begin{abstract}
The study of the dynamics of a two-body system in modified gravity constitutes a more complex 
problem  than in Newtonian gravity. Numerical
methods are typically needed to solve the equations of geodesics. Despite the complexity of the problem, 
the study of a two-body system in $f(R)$ gravity leads to  a new exciting perspective
hinting the right strategy to adopt in order to probe modified gravity. 
Our results point out some differences between the 
{\em semiclassical} (Newtonian) approach, and the {\em relativistic} (geodesic) one thus suggesting that the latter represents the 
best strategy for future tests of modified theories of gravity. { Finally, we have also highlighted the capability of forthcoming observations to serve as smoking gun of modified gravity revealing a departure from GR or further reducing the parameter space of $f(R)$ gravity}. 
\end{abstract}

\section{Introduction}
\label{intro}

The perihelion advance of Mercury is undoubtedly one of the pillars of General Relativity. 
It played a marginal role in the validation of the theory. Anecdotally, the point of view that was held as valid for a long time was to relate
the anomalous perihelion advance to the existence of the planet Vulcan, proposed in the 1860s by Le Verrier,  
lying between the Sun and Mercury which was hoped to be discovered at some point. Although the search of this planet continued till the late 1920s,
no evidences  were ever found, and it was eventually replaced by the explanation for the  perihelion advance of Mercury which is offered by converted General Relativity. Not surprisingly, this  observational probe became a solid argument to
certify the effectiveness and  validity of the theory \cite{Will2014,Will2015}. 

Nowadays, General Relativity is itself one of the fundamental pillars of the standard cosmological models.
Nevertheless, in the last decades amazing technological progresses have allowed us 
obtain sufficiently good data to probe our models of the cosmos, and the results have been quite shocking.
We seem to live in a Universe dominated by two (kind of Olympic) substances: dark matter and dark energy. But
the  lack of knowledge of their fundamental nature (whether particles or scalar fields)  \cite{Bertone2005,Feng2010,Capolupo2016,Capolupo2017,idm2017,idm201b}, or whatever physical mechanism that may cunningly mimic them remains to the present day.
This has led some sectors of the community to argue along the second possibility, which is that 
such manifestations could indicate a breakdown of General Relativity at certain scales.
After all, for many decades we were convinced that the explanation of Mercury's perihelion advance rested with Vulcan, 
while the (eventual sound) solution to the puzzle came as a modified theory of gravity ({\em i.e.} the General Relativity) as compared to the up to the date prevailing one (Newtons). Similarly, many models which 
modify/extend General Relativity have been proposed to overcome the (arguably somewhat uncomfortable) need to introduce fluid/particle realizations of dark matter and dark energy 
\cite{Nojiri:2006ri,darkmetric,Nojiri2011,PhysRept,idm2015,Nojiri2017,Sotiriou2010,FaraoniMe,Olmo2011,Cai2016,Ester1,Ester2,Ester3,Ester4,idm2018}.

The simplest route along modifications is represented by $f(R)$ gravity, which replaces the Einstein-Hilbert action, which is linear in the Ricci scalar $R$, 
with a more general function $f(R)$. In its weak field limit, $f(R)$ gravity exhibits a Yukawa-like correction term to the 
standard Newtonian potential, thus modifying the dynamics of the two body problem \cite{Annalen} as a matter of fact. But extreme care is required so that this task can be accomplished competently. On the one hand, 
any alternative relativistic theory of gravity should allow recovering General Relativity at small scales so as to match the unavoidably tight observational 
constraints  \cite{Will2014}. On the other hand, two-particle dynamics in the context of the modification of the Newtonian 
potential offers a straightforward possibility to probe the gravitational field.
As it is well known, a Yukawa modification of the Newtonian potential naturally arises in the context of theories that attempt to unify 
gravity with other fundamental forces \cite{Gibbons1981}. Relatedly, it provides a possible route to explain the Pioneer anomaly 
\cite{Anderson1998,Anderson2002} and propounds predictions of the capability of forthcoming experiments to constrain departures from General Relativity \cite{Hees2017}. 

A promising workbench to test the underlying theory of gravity is the dynamics of binary systems composed 
by either two pulsars \cite{deLa_deMa2014,deLa_deMa2015,DeLaurentis2012,Berti,Antoniadis,Freire,Shao}, or a pulsar and a supermassive black hole (SMBH) \cite{Liu2012,DeLaurentis2016}. 
Although systems like the latter have not been observed yet, prospects of a future detection have increased enormously thanks to the 
 BlackHoleCam\footnote{https://BlackHoleCamorg} project
and Event Horizon Telescope (EHT)\footnote{http://www.eventhorizontelescope.org} survey that is 
mapping the central region 
of the Milky Way around the SMBH Sagittarius A* (Sgr A*) \cite{Goddi2017,Mizuno2018}. It thus becomes a matter of quite relevance to develop improved descriptions of two body system dynamics in modified gravity. Our attention here focuses on possible
methodologies within the $f(R)$ setting. 

In Sect. 2 we summarize the general framework and the previous results. In the Sect. 3 we outline how to perform the integration of the equation of motion, and we particularize the solutions to the S0-2 star orbiting around Sgr A*. In Sect. 3, we show ans discuss our results. Finally, in Sect. 4, we give our conclusions.

\section{$f(R)$-Yukawa like potential}
As mentioned above, in its weak field limit, $f(R)$ gravity modifies the Newtonian potential.
The functional form of this modification is the following:
\begin{equation}\label{eq:potyuk}
 \Phi(r) = -\frac{G M}{(1+\delta)r}(1+\delta e^{-r/\lambda}).
\end{equation}
Here $r$ is the distance of a test particle from the source of the gravitational field, $G$ is Newton's constant, and
$\delta$ and $\lambda$ are the strength of the Yukawa correction and the scale over which  the latter acts, respectively.

The scale $\lambda$ deserves a deeper discussion. Although it is usually identified as the Compton length of a massive graviton 
$\lambda_g = hc/m_g\approx 10^4$ AU  with $m_g\sim 10^{-22}$ eV \cite{Lee2010,Abbott2017}, the parameter $\lambda$ in Eq. 
\eqref{eq:potyuk} arises from the extra degrees of freedom of $f(R)$ gravity. In general, 
it has been demonstrated that a $(2k+2)$-order theory  gives rise to $k$ extra gravitational scales \cite{Quandt1991}. Therefore, 
it can be of the same order of magnitude of the $\lambda_g$, but it can also assume different values depending on the particular scale under investigation (e.g. at the scales of galaxy clusters $\lambda\sim$ Mpc \cite{Cap-def-Sal2009,demartino2014,demartino2016}). 
Thus, to test departures from the Newtonian potential, one can follow two prescriptions: 
the first is a sort of {\em semiclassical} approach in which  
the modified gravitational potential in Eq. \eqref{eq:potyuk} is 
introduced
in the Newtonian equations of motion \cite{Borka2012,Borka2013,Zakharov2016,idmRLmdl2017,Zakharov2018}; 
the second is a full {\em relativistic} approach where the equations of the geodesics are taken into account. 

In the {\em semiclassical} approach, the equation of motion of a massive point-like particle ($m$) moving in the modified gravitational potential well generated by the particle $M$ arises as solution of
\begin{align}
\label{eq:1}  \ddot{r} = -\nabla\Phi(r)\,. 
\end{align}
In Newtonian mechanics no orbital precession happens. Nevertheless, the Yukawa-term in the Newtonian potential gives rise to a precession of the orbit \cite{Borka2012,Borka2013,Zakharov2016,idmRLmdl2017,Zakharov2018}. 
Moreover, for those binary systems having their semi-major axis much smaller than the Yukawa length, it is possible to 
find an analytical expression to compute the precession \cite{idmRLmdl2017}:
\begin{align}\label{eq:precYuk}
\Delta \varphi \approx 2\left|\pi\sqrt{1+\frac{2 \delta }{2  (1+\delta)(\lambda/a_\epsilon)^2 -3 \delta}}\right| - 2\pi\,,
\end{align}
where $a_\epsilon\equiv a(1-\epsilon)$, and $a$ is the semi-major axis and $\epsilon$ is the eccentricity.

Nevertheless, as it has been already anticipated, the modified gravitational potential in Eq. \eqref{eq:potyuk} arises from $f(R)$ gravity. It follows from solving the fourth order field equations in the post-Newtonian limit of a spherically symmetric metric. The general solution of the problem, obtained by matching at infinity the Minkowski 
space-time, is given by \cite{Annalen}:
\begin{equation}
\label{metric0}
ds^2=\left[1+\Phi(r)\right]dt^2-\left[1-\Psi(r)\right]dr^2-r^2d\Omega\,,
\end{equation}
where the functional forms of  $\Phi(r)$ and $\Psi(r)$ are:
\begin{eqnarray}
\label{eq:PHI}\Phi(r) &=& -\frac{2G M \left(\delta  e^{-\frac{r}{\lambda}}+1\right)}{rc^2(\delta +1)}, \\
\label{eq:PSI}\Psi(r) &=& - \frac{2G M }{c^2(\delta +1)r}\biggl[1-\delta  e^{-\frac{r}{\lambda}}\left(1+
\frac{r}{\lambda}\right)\biggr]\,.
\end{eqnarray}
In such a {\em relativistic} approach, the equations of motion, {\em i.e.} the equations of geodesics, 
once written for the non-zero Levi-Civita connections, are given by \cite{idmRLmdl2017pII}
\begin{eqnarray}
{\ddot r} =&&\Delta^{-1}\biggl[R_S \left({\dot r}^2-{\dot t}^2\right) \left(\delta  (\lambda +r)+ e^{\frac{r}{\lambda }} \lambda\right)
\nonumber\\&&
+e^{\frac{r}{\lambda }} \lambda (1+\delta ) r^3 \left({\dot \theta}^2+\sin^2\theta {\dot \phi}^2\right)\biggr]\,, \label{eq:geo1}
\end{eqnarray}
\begin{eqnarray}
\label{eq:geo2} 
{\ddot \theta} =&&\cos\theta \sin\theta {\dot \phi}^2 -  \frac{2 {\dot r} {\dot\theta}}{r}\,,
\end{eqnarray}
\begin{eqnarray}
\label{eq:geo3}  {\ddot \phi} =&& -\frac{2 {\dot \phi}}{r} \left[{\dot r}+\cot\theta r {\dot \theta}\right]\,,\\[0.2cm]
\end{eqnarray}
\begin{eqnarray}
\label{eq:geo4}  {\ddot t} =&& \Delta^{-1}\biggl[2 R_S \left[\left(e^{\frac{r}{\lambda }}+\delta \right) \lambda +\delta  r\right] {\dot r} {\dot t}\biggr]\,,
\end{eqnarray}
where, for the sake of convenience, we have defined
\begin{equation}
\Delta\equiv\lambda  r \left[2 R_S \delta +e^{\frac{r}{\lambda }} \left(2 R_S- (1+\delta ) r\right)\right]\,,
\end{equation}
and
\begin{equation}
R_S = \frac{GM}{c^2}\,.
\end{equation}
As it was the case for the  {\em semiclassical} approach, the numerical integration of the geodesic equation  shows the precession of the orbit, and it is possible to obtain an analytical expression of it for those systems having their semi-major axis much smaller than the Yukawa length \cite{idmRLmdl2017pII}:
 \begin{eqnarray}\label{eq:deltaphi}   
\Delta\phi\approx \frac{\Delta\phi_{GR} }{(\delta +1)}\,,
  \end{eqnarray}
where $\Delta\phi_{GR} $ is the general relativistic precession  given by
\begin{equation}\label{eq:precGR}
 \Delta\phi_{GR}= \frac{6 \pi  G M}{a c^2 \left(1-\epsilon^2\right)}\,.
\end{equation}

To understand the differences between two approaches, and to investigate the impact of the parameter $\lambda$, one should integrate the equation of motion for a physical system and analyse the impact of both approaches on the resulting orbits.

\section{Numerical integration: results and discussions} 

\begin{figure*}[!ht]
\centering
\hspace*{-1.0cm}
\includegraphics[width=1.9\columnwidth]{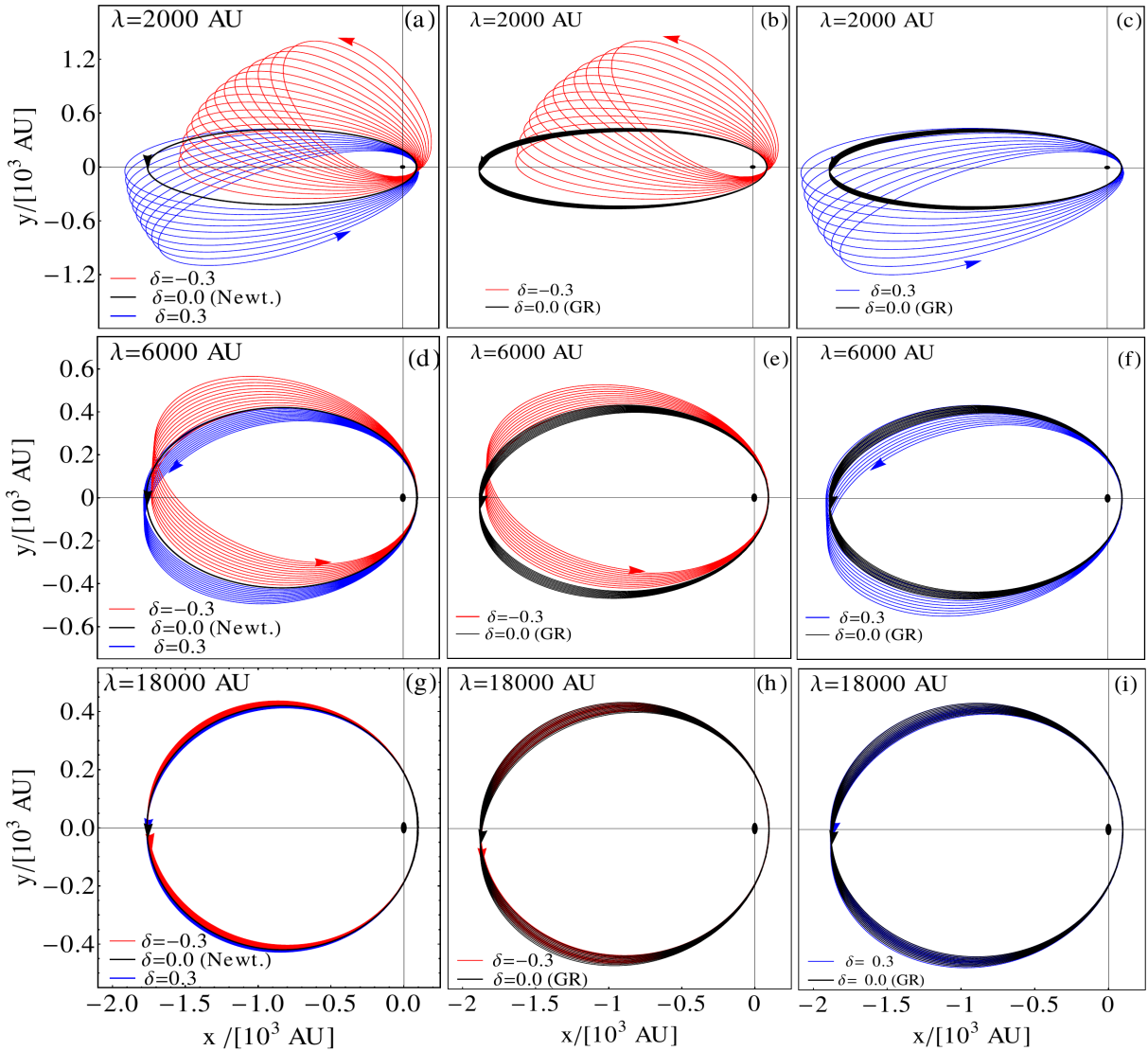}
\caption{Numerical integration of the equation of motion. Specifically, panels (a), (d), and (g) show the results of the 
{\em semiclassical} approach, while in the other panels the results of the numerical integration of the geodesics are depicted.}\label{fig1}
\end{figure*}

We have integrated numerically the differential equations of motion for the {\em semiclassical approach}, Eq. \eqref{eq:1}, and for the {\em relativistic} one, Eqs. \eqref{eq:geo1}-\eqref{eq:geo4}.  The problem is particularized for the orbits of the S0-2 star around Sgr A*. Following \cite{Borka2013}, we fixed the mass of central object to $M= 4.3\times 10^6 M_\odot$, while the initial conditions of the orbital parameters are taken from \cite{Gillessen2009}. 
In Fig. \ref{fig1}  we present the results of the integration to show the impact of both parameters $(\delta, \lambda)$ on the orbital motion.  We have fixed the parameter $\delta$ to the values $[-1/3,\, 0,\, 1/3]$ that are represented with red, black and blue lines, respectively. In the left column, the results of the {\em semiclassical} approach are shown, while in the 
central and right  the results of the {\em relativistic} approach for $\delta=-1/3$ and $\delta=1/3$ are shown respectively.
The value of the scale length is fixed (from  top to  bottom) to $[2,\, 6,\, 18]\times10^3$ AU. 
From these results one can argue the importance of using geodesics instead of the classical mechanics equations of motion. Indeed, as long as the scale length is of the order 
of the semi-major axis (i.e. the top and the central rows) both approaches lead to similar results. 
The main differences is that the {\em relativistic} approach predicts a slightly larger or smaller precession in the 
cases of $\delta>0$ or $\delta<0$, respectively. The issue turns out when $\lambda\gg r$. In such a case, 
as illustrated in the bottom row, the predictions of both approaches tend to be reduced to their Newtonian and general relativistic limit. 
Thus, while the {\em semiclassical} approach would predict zero orbital precession, the
{\em relativistic} one would correctly predict $\Delta\phi_{GR}$. This can be easily understood looking at the potentials $\Phi(r)$ 
and $\Psi(r)$. In case $\lambda\gg r$, one finds:
\begin{eqnarray}
\Phi(r) \longrightarrow -\frac{2G M }{rc^2}\,, \qquad
\Psi(r) \longrightarrow - \frac{2G M (1-\delta)}{c^2(1+\delta )r}\,.
\end{eqnarray}
Therefore, the {\em semiclassical} approach reduces to the Newtonian setting, while the 
{\em relativistic} approach reduces to the General Relativity one plus a correction that still remains in $\Psi(r)$.

\begin{figure}[!ht]
	\centering
	\hspace*{-1.0cm}
	\includegraphics[width=1.0\columnwidth]{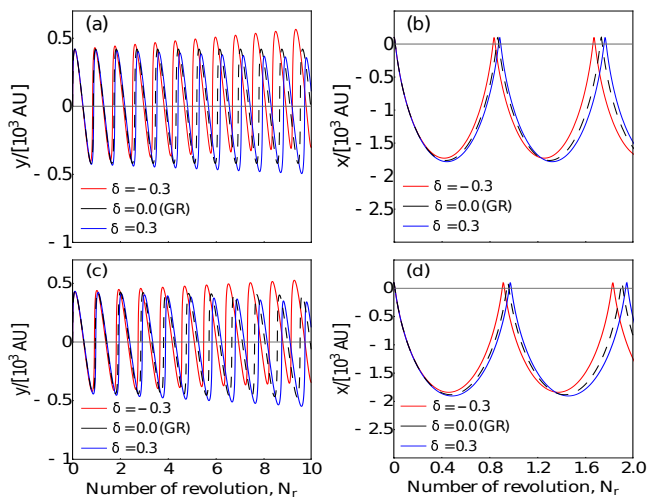}
	\caption{Equation of motion as function of the number of revolutions. Panels (a), (b)  show the results of the 
		{\em semiclassical} approach, while panels (c) and (d) show the numerical integration of the geodesics. The plots are particularized for the case $\lambda=6000$ AU. }\label{fig2}
\end{figure}

To quantify the variation of the orbital motion between the two approaches, we have focused on the case $\lambda=6000$ AU represented by panels (d) to (f) of Fig. \ref{fig1}. Thus, in Fig. \ref{fig2}, we have depicted the two spatial coordinates as a function of the number of revolutions ($N_r$). The panels (a) and (b) show the {\em semiclassical} results corresponding to panel (d) of Fig. \ref{fig1}, while the panels (c) and (d) depict the results 
of the {\em relativistic} approach and correspond to panels (e) and (f) in Fig. \ref{fig1}. The variations of the orbits in the Yukawa potential are clearly viewable. In both cases, we can point out a small enhancement/reduction of the semi major axis, {\em i.e.} the $x$ coordinate, and a larger variation of $y$. Thus, it is straightforward to define the relative variation for both coordinates:
\begin{align}
& \Delta x_{\delta_{\pm}} = \frac{x(t) - x_{GR}(t)}{x(t)}\,,\\
& \Delta y_{\delta_{\pm}} = \frac{y(t) - y_{GR}(t)}{y(t)}\,,
\end{align}
where $\delta_{\pm}$ indicates the positive or negative values of  $\delta$, $x(t)$ and $y(t)$ are the solutions of the equation of motions in the Yukawa-like gravitational potential, and $x_{GR}$ and $y_{GR}$ are the solution of the equation of motion in the GR.
All results are summarized in Table \ref{tab:1}.
Thus, for example, we find that the relative difference in the $x$-direction is $\sim 1\%$ for $\delta=-1/3$, and it is constant in time. While, in the $y$-direction, we find $\Delta y \sim 3\%$ after one revolution, which becomes  $\sim 15\%$. The differences between the two approaches are at order of $2\%$ in the  $y$-direction after five revolutions, while in the $x$-direction are totally negligible. These differences with GR can be easily translated in angular deviations on the sky. Since the distance from the Sgr A* is known, we find $\Delta x \sim 1\% \sim 0.002^{\prime\prime}$ and $\Delta y \sim 3\% \sim 0.003^{\prime\prime}$ (which becomes $\sim 0.01^{\prime\prime}$ after five revolutions). From one hand, we have to mention that a measurement with an accuracy of milliarcseconds is within the capability of current and forthcoming interferometers which are tracking (or will trace) the orbits of the S-stars around the galactic center \cite{Gillessen2009,Gillessen2017}, {and which will constrain the pericenter advance at level of $\mu$as within few years \cite{Grould2017}}.
On the other hand, we should mention that the value of the parameter $\delta$ could be much lower implying a huge reduction of the difference with the general relativistic predictions and, as matter of fact, hiding the modification of gravity. Finally, the difference between the two approaches becomes relevant for the orbital motion of the S-stars when the value of $|\delta|>1/3$, and/or the value of $\lambda$ is much lower than 6000 AU, while for higher values any modification becomes negligible.

\begin{table}[!h]
	\begin{center}
		\caption{Relative variation of the spatial coordinate with respect to the general relativistic value, after one and five revolution. The subscripts $\delta_{\pm}$ indicates the two values that have been considered $\pm0.3$, while $\lambda$ is fixed to $6000$ AU.}
		\label{tab:1}
		\hspace*{1.0cm} {\bf{\em semiclassical}} \hspace*{1.0cm} {\bf{\em relativistic}}
		\begin{tabular}{|l|cc||cc|}
			\hline
			& $N_r = 1$ & $N_r = 5$ & ${N}_r = 1$ & ${N}_r = 5$\\
			\hline
			$\Delta x_{\delta_{-}}$  & 2.31\% & 2.43\% & 2.26\% & 2.33\%\\
			$\Delta x_{\delta_{+}}$  & 1.01\% &1.03\% & 1.01\%  & 1.02\% \\
			$\Delta y_{\delta_{-}}$ & 3.32\% & 14.98\%  & 3.21\% & 13.71\% \\
			$\Delta y_{\delta_{+}}$  & 2.01\% & 9.46\% & 1.76\% & 8.24\%\\
			\hline
		\end{tabular}
	\end{center}
\end{table}

It is worth to note that others analysis comparing the predicted orbital motion of S0-2 star around the Galactic Centre and data have been carried out in the framework of both $f(R)$ and hybrid gravity \cite{Capozziello2014,Capozziello2016}. In both cases, the orbits are obtained integrating the Newtonian equation of motion with a modified gravitational potential according to the specific theory. For both theories, the results pointed out very small departures from the general relativistic case. Moreover, these departures are of the same order of what we have predicted with our approach confirming once again our results.

\section{Frame related issues}

At this point a further discussion on the results is required. All calculations have been performed in the Jordan frame, where a self-interacting potential is taken into account, but the same calculations can also be carried out in the Einstein frame.  In the latter case, the gravitational coupling becomes a function of space and time and depends  dynamically from
scalar fields.  As a consequence, according to the Mach Principle, 
the gravitational interaction changes with distance and time \cite{Stabile1,Stabile2}. 
Although  a conformal transformation connects the two frames, it is still an opened question whether these two frames are mathematically, or even physically, equivalent. For example, let us consider geodesic motion: in the Jordan frame, a massive test  particle falls along time-like while in the Einstein frame it does not due to a force arising from the gradient of the conformal scalar field. Therefore, the Equivalence  Principle only holds in the Jordan frame. 
Nevertheless, although gauge invariance is broken, there is the possibility, when computing the weak field limit, to check step by step the perturbative approach in both frames to obtain self-consistent results allowing to compare potentials, masses and other physical quantities.

\section{Conclusions}

We have devoted this analysis to point out the importance of using geodesics to test modified  gravity with binary systems, and to the possibility of revealing the modification of gravity along the orbital path of an S-star at the Galactic center. We have summarized
the results presented in \cite{idmRLmdl2017,idmRLmdl2017pII}, and then we have numerically integrated the equations of motion particularizing our analysis to the S0-2 star. In Fig. \ref{fig1}, we have shown the impact of both parameters of the Yukawa-like gravitational potential on the orbital motion around Sgr A*. The modification of the gravitational potential is reflected with different magnitude in both spatial coordinates. While the variation with respect the general relativistic predictions is almost negligible in the $x$-direction, it is not in the $y$-direction. In fact, after five revolutions $y$, as predicted in the Yukawa-like potential, differs form the GR one $\sim 0.01^{\prime\prime}$. { The level of accuracy required to reveal modification of gravity along the geodesic track of the S0-2 star should be reached in few years \cite{Gillessen2017,Grould2017} serving as a smoking gun of modified gravity.

Despite the intrinsic difficulties of the methodology applied in this analysis, that are mainly related to the boundary conditions problem, and the astrophysical limits of current and forthcoming datasets, we should point out that the same methodology can be also applied to closer binary system, such as neutron stars binary, which also have a smaller orbital period, to take advantage of the very precise timing measurements of pulsars and  improving the current constraints on modified gravity models.}

\section*{Acknowledgements}
 
This article is based upon work from COST Action CA1511 Cosmology and Astrophysics 
Network for Theoretical Advances and Training Actions (CANTATA), 
supported by COST (European Cooperation in Science and Technology).
MDL  acknowledges the INFN Sez. di
Napoli (Iniziative Specifiche QGSKY and TEONGRAV). MDL is supported by the Grant "BlackHoleCam"
Imaging the Event Horizon of Black Holes awarded by the
ERC in 2013 (Grant No. 610058).


\begin{thebibliography}{99}

\bibitem{Will2014}
C.M. Will, Living Rev. Relativ., 17, 4 (2014)

\bibitem{Will2015}
C.M. Will Class. Quantum Grav., 32, 124001 (2015)

\bibitem{Bertone2005}
G. Bertone, D. Hooper, J. Silk,  Phys. Rept.,  405, 279 (2005)

\bibitem{Feng2010}
J.L. Feng, Ann. Rev. Astron. Astrophys., {48}, 495 (2010) 



\bibitem{Capolupo2016}
A. Capolupo, 
{Advances in High Energy Physics}, { 2016}, 8089142, 10 (2016).

\bibitem{Capolupo2017}
A. Capolupo,  
{ Advances in High Energy Physics}, 2018, 9840351, 7 (2018).

\bibitem{idm2017} 
 I. De Martino, T. Broadhurst, H. Tye, T. Chiueh, Hsi-Yu Schive, R. Lazkoz, Phys. Rev. Lett, 119, 22110 (2017)
 
 \bibitem{idm201b} 
 I. De Martino, T. Broadhurst, H. Tye, T. Chiueh, Hsi-Yu Schive,  arXiv:1807.08153 (2018)

\bibitem{Nojiri:2006ri}
S.~Nojiri and S.~D.~Odintsov,
Int.\ J.\ Geom.\ Meth.\ Mod.\ Phys.\,  4, 115 (2007)


\bibitem{darkmetric}
S. Capozziello, M. De Laurentis, M. Francaviglia, S. Mercadante, 
Foundations of Physics, { 39}, 1161 (2009).

\bibitem{Nojiri2011}
S. Nojiri, S. D. Odintsov, Phys.Rept., 505,  59-144 (2011)

\bibitem{PhysRept}
S. Capozziello, M. De Laurentis,  Phys. Rept., 509, 167 (2011).

\bibitem{idm2015}
I. de Martino, M. De Laurentis, S. Capozziello,  Universe, 1, 123 (2015)

\bibitem{Nojiri2017}
S. Nojiri, S. D. Odintsov, V.K. Oikonomou, 
Phys.Rept., { 692}, 1-104 (2017)

\bibitem{Sotiriou2010}T.P. Sotiriou, V. Faraoni, Rev. Modern Phys. 82 , 451 (2010).

\bibitem{FaraoniMe}S. Capozziello, M. De Laurentis, and V. Faraoni, Open Astr.
J. 2, 1874 (2009).

\bibitem{Olmo2011}G. J. Olmo, Int. J. Mod. Phys. D 20, 413 (2011).

\bibitem{Cai2016}Y.-F. Cai, S. Capozziello, M. De Laurentis, and E. N.
Saridakis, Rep. Prog. Phys. 79, 106901 (2016).
\bibitem{Ester1}S. Capozziello, E. Piedipalumbo, C. Rubano, P. Scudellaro 	Astron.Astrophys. 505, 21, (2009)
\bibitem{Ester2}S. Capozziello, P.K.S. Dunsby, E. Piedipalumbo, C. Rubano, Astron.Astrophys. 472, 51 (2007)
\bibitem{Ester3} M.Demianski, E. Piedipalumbo, C.Rubano, P. Scudellaro, Astron.Astrophys. 481, 279 (2008)
\bibitem{Ester4}M. Demianski,  E. Piedipalumbo, C. Rubano, C. Tortora, Astron.Astrophys.  454, 55 (2006)

\bibitem{idm2018} 
I. de Martino, Symmetry 10, 372 (2018)
\bibitem{Annalen} 

S. Capozziello, M. De Laurentis, Annalen der Physik { 524}, 545 (2012).
\bibitem{Gibbons1981}
G. W. Gibbons, B. F. Whiting, Nature, 291, 636-638 (1981)

\bibitem{Anderson1998}
J. D. Anderson, P. A. Laing, E. L. Lau, A. S. Liu, M. M. Nieto, and S. G. Turyshev, 
\prl 81, 2858 (1998)

\bibitem{Anderson2002}
J. D. Anderson, P. A. Laing, E. L. Lau, A. S. Liu, M. M. Nieto, and S. G. Turyshev, 
\prd \, 65, 082004 (2002)

\bibitem{Hees2017}
A. Hees, T. Do, A. M. Ghez, G. D. Martinez, S. Naoz, E. E. Becklin, A. Boehle, S. Chappell, 
D. Chu, A. Dehghanfar, et al.
\prl 118, 211101, (2017)

\bibitem{deLa_deMa2014}
M. De Laurentis, I. de Martino, \mnras { 431}, 1, 741 (2013)

\bibitem{deLa_deMa2015}
M. De Laurentis, I. de Martino, Int. J. Geom. Math. Meth., 1250040D (2015)

\bibitem{DeLaurentis2012}M. De Laurentis, R. De Rosa, F. Garufi, L. Milano, Mon. Not.
R. Astron. Soc. 424, 2371 (2012)

\bibitem{Berti}
E. Berti, E. Barausse, V. Cardoso, L. Gualtieri, P. Pani, et al.	
Classical and Quantum Gravity, { 32}, 24,  243001 (2015).

\bibitem{Antoniadis}J. Antoniadis, Astrophysics and Space Science Proceedings,
40,1 (2015)

\bibitem{Freire} P. C. C. Freire, N. Wex, G. Esposito-Farèse et al. MNRAS 423
3328, (2012)

\bibitem{Shao} L. Shao, N. Sennett, A. Buonanno, M. Kramer, N. Wex,
arXiv:1704.07561, (2017)

\bibitem{Liu2012}
K. Liu, N. Wex, M. Kramer, J. M. Cordes, and T. J. W. Lazio, \apj { 747},1 (2012).

\bibitem{DeLaurentis2016}M. De Laurentis, Z. Younsi, O. Porth, Y. Mizuno, L. Rezzolla Phys. Rev. D 97, 104024 (2018).

\bibitem{Goddi2017}
C. Goddi, H. Falcke, M. Kramer, L. Rezzolla, et al., 2017, Modern Physics D  { 26}, 02 (2017)

\bibitem{Mizuno2018}Y. Mizuno, Z. Younsi, C. M. Fromm, O. Porth, M. De Laurentis, H. Olivares, H. Falcke, M. Kramer, L. Rezzolla Nature Astronomy, 2, 585 (2018)

\bibitem{Lee2010}
K. Lee, F. A. Jenet, R. H. Price, N. Wex, M. Kramer
\apj \, 722, 1589-1597 (2010) 
 
\bibitem{Abbott2017}
B. P. Abbott et al. (LIGO Scientific and Virgo Collaboration)
\prl \, 118, 221101 (2017)

\bibitem{Quandt1991}
 I. Quandt, H. J. Schmidt, Astron. Nachr., 312, 97 (1991)
 
\bibitem{Cap-def-Sal2009}
S. Capozziello, E. De Filippis, V.Salzano,  \mnras { 394}, 947 (2009)
 
\bibitem{demartino2014}
I. de Martino, M. De Laurentis, F. Atrio-Barandela, S. Capozziello,
\mnras \, 442, 2, 921-928 (2014)

\bibitem{demartino2016}
I. de Martino, \prd \,  93, 124043 (2016) 

\bibitem{Borka2012}
D. Borka, P. Jovanovi\'c, V. Borka Jovanovi\'c and A. F. Zakharov,
\prd { 85} 124004 (2012)

\bibitem{Borka2013}
D. Borka, P. Jovanovi\'c, V. Borka Jovanovi\'c, A. F. Zakharov,
\jcap { 11} 050 (2013)

\bibitem{Zakharov2016}
A.F. Zakharov and P. Jovanovi\'c and D. Borka and V. Borka Jovanovi\'c, \jcap { 05}, 045, (2016)

\bibitem{idmRLmdl2017}
I. de Martino, R. Lazkoz, M. De Laurentis, 
\prd, 97, 104067 (2018)

\bibitem{Zakharov2018}
A. F. Zakharov, P. Jovanovi\'{c}, D. Borka, V. Borka Jovanovi\'{c},   	arXiv:1801.04679 (2018) 

\bibitem{idmRLmdl2017pII}
M. De Laurentis, I. De Martino, R. Lazkoz, \prd 
97, 104068 (2018)

\bibitem{Gillessen2009}
  S. Gillessen, F. Eisenhauer, T. Fritz, H. Bartko, K. Dodds-Eden et al.,
\apj 707  L114 (2009)

\bibitem{Gillessen2017}
S. Gillessen, P. Plewa, F. Eisenhauer et al. \apj 837,  1 (2017).

\bibitem{Grould2017}
M. Grould, F. H. Vincent, T. Paumard, G. Perrin, 
\aea, 608, A60 (2017)

\bibitem{Stabile1}
A. Stabile Phys. Rev. D 82, 064021 (2010)

\bibitem{Stabile2}
A. Stabile, An. Stabile, S. Capozziello Phys. Rev. D 88, 124011 (2013)

 \bibitem{Younsi2016}
Z. Younsi, A. Zhidenko, L. Rezzolla, R. Konoplya, and Y. Mizuno, \prd 94, 084025 (2016)


\bibitem{Younsi2017}
 Z. Younsi and A. Grenzebach, in Proceedings, 14th Marcel Grossmann Meeting on Recent Developments in Theoretical and Experimental General Relativity, Astrophysics, and Rela- tivistic Field Theories (MG14) (In 4 Volumes): Rome, Italy, July 12-18, 2015, Vol. 4 (2017) pp. 3525:3530.
 
\bibitem{Capozziello2014}
S. Capozziello, D. Borka, P. Jovanovi\'{c}, V. Borka Jovanovi\'{c}
\prd, 90, 044052  (2014)

\bibitem{Capozziello2016}
D. Borka, S. Capozziello, P. Jovanovi\'{c}, V. Borka Jovanovi\'{c}
Astropart. Phys. 79, 41  (2016) 
 
\end{thebibliography}
\end{document}